# Towards a new approach of continuous process improvement based on CMMI and PMBOK

Yassine Rdiouat[1], Naima Nakabi[2], Khadija Kahtani[3] and Alami Semma[4]

[1] Department of Mathematics and Computer Science, Faculty of Science and Techniques, Hassan 1 University, Settat, Morocco

[2] Department of Mathematics and Computer Science, Faculty of Science and Techniques, Hassan 1 University, Settat, Morocco

[3] Department of Mathematics and Computer Science, Faculty of Science and Techniques, Hassan 1 University, Settat, Morocco

[4] Department of Mathematics and Computer Science, Faculty of Science and Techniques, Hassan 1 University, Settat, Morocco

**Abstract**

A process-centric approach helps an organization to improve the way it works with. It allows scalability and provides a way to capitalize knowledge on best practices. It also makes better use of resources and helps to understand trends.

PMBOK is a project management methodology, while CMMI is a model for process improvement.

In this paper, we conduct a study on PMBOK and CMMI frameworks to show that they can be converged and complementary.

We expect this paper research will be useful for organizations to deploy a new approach of continuous process improvement based on pooling CMMI and PMBOK.

## 1. Introduction

Nowadays, it is very rare that the companies develop alone all the elements that make up a product or service. Most of the time, some components are created in-house while others are acquired. All these components are then integrated in the final product or service. Thus, the organizations must manage and control their complex development and maintenance processes.

Problems faced by organizations involve solutions for the whole company and require an integrated approach to their development activities to reach their strategic objectives.

Generally, there are three critical dimensions on which rely all organizations to increase their effectiveness: Skills and human potential, Procedures and methods, Technology and technical expertise. But, what is the unifying element which keeps all together? It is the processes used by the organization that are the deepest basis, the more sustainable, which supports all of these dimensions. Many organizations are not interested to adopt process improvement approach because they focus only to have great staff and technology (HERO Model). These organizations encounter serious quality problems since the system quality depends heavily on the processes quality of the organization.

We do not claim that people and technology are not important. Technology is constantly changing and people change jobs all the time. A process-centric approach provides infrastructure to face these changes and to maximize the productivity of individuals and the use of technology to be more competitive.

PMBOK is a standard which presents techniques and tools for effective project management, while, CMMI is a processes maturity model that provides the essential elements of effective processes; it focuses on the organization maturity and not on individuals.

For an organization which adopts PMBOK and wants to improve continuously their processes, CMMI is a good candidate that provides more detailed roadmap for process improvement.

The objective of this paper research is not to describe how to deploy CMMI, but to show how CMMI can provides added value to institutionalize repeatable, effective, efficient and scalable processes.

In this paper we attempt to:
- Make a map of PMBOK (4th ed.) and CMMI v1.3 models processes.
- Detect similarities between the two frameworks.
- Present how CMMI can supplement PMBOK and how PMBOK can complement CMMI with more tools and guidance.

## 2. Methodology

In order to conduct a comparative study effectively, we estimate necessary to make a mapping between the two frameworks.

As we know, it is always difficult to know the appropriate granularity to perform a mapping. On one hand, a mapping high level may not provide enough information on the similarities and differences. On the other hand, a mapping of very low level can generate a large number of connections between the two models making very complex correspondence. The study we have conducted is a compromise between the two levels. In this paper, we try to have, first, a high level comparison to understand differences between the two models architectures, then, we attempt to map CMMI process areas (PAs) to PMBOK processes in order to have more details and a better understanding of the differences and similarities between the both frameworks.

## 3. Definitions

### 3.1 CMMI

CMMI (Capability Maturity Model Integration) is a process improvement approach that provides organizations with the essential elements of effective processes, which will improve their performance [1]. It includes three structured models: Development, Acquisition and Services. For our study, we are interested to CMMI for development. CMMI for Development is a collection of best practices relating to the development's activities and maintenance. It provides a comprehensive integrated set of guidelines for developing products and services [2]. It involves practices that cover the product lifecycle from conception to delivery and maintenance. It focuses on the work necessary to build and maintain the entire product [3].

CMMI provides two representations staged and continuous. In the staged representation, maturity level of an organization ranges from level 1 to 5. In the continuous representation, each process capability level ranges from 0 to 5. The staged representation is most suitable for an organization that does not know which processes need to be improved first; it offers process areas applicable to each maturity level. However, the continuous representation provides flexibility to select processes that fits for achieving business goal of the organization [4].

The table 1 compares the six capability levels to the five maturity levels. The names of the four higher levels are the same in both representations. The differences are basically in the first levels. There is no level 0 in the staged representation and, the level 1, is called "Performed" while the maturity level is called "Initial". So, the starting point is not the same in both representations [3].

Table 1: Comparison of CMMI capability and maturity levels

| Level | Continuous representation Capability level | Staged representation Maturity level |
|---|---|---|
| 0 | Incomplete | N/A |
| 1 | Performed | Initial |
| 2 | Managed | Managed |
| 3 | Defined | Defined |
| 4 | Quantitative Managed | Quantitative Managed |
| 5 | Optimizing | Optimizing |

- **Capability Level 0 – Incomplete**
  An "incomplete process" is a process that is either not performed or partially performed [3]. One or more of the specific goals of the process area are not satisfied and no generic goals exist for this level since there is no reason to institutionalize a partially performed process.

- **Capability Level 1 – Performed**
  The process is unstructured (ad hoc). Few activities are explicitly defined and success depends on individual effort and heroics. Organizations often produce products and services that work; however, they frequently exceed the budget and schedule of their projects. Organizations are characterized by a tendency to over commit, abandon processes in the time of crisis, and not be able to repeat their past successes [3]

- **Capability Level 2 - Managed**
  Basic project management processes are established to track cost, schedule, and functionality. The necessary discipline is in place to repeat earlier successes.
  In other words, the projects of the organization have ensured that requirements are managed and that processes are planned, performed, measured, and controlled.
  The process discipline reflected by maturity level 2 helps to ensure that existing practices are retained during times of stress. When these practices are in place, projects are performed and managed according to their documented plans [3].

- **Capability Level 3 - Defined**
  Processes are well characterized and understood, and are described in standards, procedures, tools, and methods.

A critical distinction between maturity level 2 and maturity level 3 is the scope of standards, process descriptions, and procedures. At maturity level 2, the standards, process descriptions, and procedures may be quite different in each specific instance of the process (for example, on a particular project). At maturity level 3, the standards, process descriptions, and procedures for a project are tailored from the organization's set of standard processes to suit a particular project or organizational unit. The organization's set of standard processes includes the processes addressed at maturity level 2 and maturity level 3. As a result, the processes that are performed across the organization are consistent except for the differences allowed by the tailoring guidelines [3].

- **Capability Level 4 - Quantitative Managed**
  Detailed measures of the process and product quality are collected. Both the process and products are quantitatively understood and controlled. Quantitative objectives for quality and process performance are established and used as criteria in managing processes. Quantitative objectives are based on the needs of the customer, end users, organization, and process implementers. Quality and process performances are understood in statistical terms and are managed throughout the life of the processes [3].

- **Capability Level 5 – Optimizing**
  An optimizing process is quantitatively managed (capability level 4), it is ameliorated by understanding the common causes of process variations [3]. Continuous process improvement is enabled by quantitative feedback for the process and from piloting innovative new ideals and technologies.
  Capability Level 4 focuses on establishing baselines, models, and measurements for process performance. Capability Level 5 focuses on studying performance results across the organization or entire enterprise, finding common causes of problems in how the work is done (the processes used), and fixing the problems in the process. The fix would include updating the process documentation and training involved where the errors were injected.

The main component of CMMI is the process area. CMMI for development provides 22 process areas [5] organized into 4 categories (Project Management, Support, Engineering, and Process Management) and 5 maturity levels. The following table illustrates clearly this categorization:

Table 2: Process Areas categorization for staged representation

| Category | Process Areas | | Maturity Level |
|---|---|---|---|
| Process Management | Organizational Process Focus | OPF | 3 |
| | Organizational Process Definition | OPD | 3 |
| | Organizational Training | OT | 3 |
| | Organizational Process Performance | OPP | 4 |
| | Organizational Innovation and Deployment | OID | 5 |
| Support | Configuration Management | CM | 2 |
| | Process and Product Quality Assurance | PPQA | 2 |
| | Measurement and Analysis | MA | 2 |
| | Decision Analysis and Resolution | DAR | 3 |
| | Causal Analysis and Resolution | CAR | 5 |
| Engineering | Requirements Development | RD | 3 |
| | Technical Solution | TS | 3 |
| | Product Integration | PI | 3 |
| | Verification | VER | 3 |
| | Validation | VAL | 3 |
| Project Management | Requirements Management | REQM | 2 |
| | Project Planning | PP | 2 |
| | Project Monitoring and Control | PMC | 2 |
| | Supplier Agreement Management | SAM | 2 |
| | Integrated Project Management | IPM | 3 |
| | Risk Management | RSKM | 3 |
| | Quantitative Project Management | QPM | 4 |

To gain a better understanding of the focus of CMMI, we shall consider the categories described above. If we were going to translate the categories into everyday business process language, we would suggest the following [6]:

- Process Management: Refers to Strategic Process Management Planning. In one of its few efforts to cross reference, CMMI suggests, that the Process Area: Organizational Process Definition, is sometimes called "process architecture."

- Support: Refers to what managers do when they manage business processes. They Plan, Organize, Monitor and Control the work of those actually implementing the process.

- Engineering: Refers to the analysis and design of new hardware or software products. In essence, one would say this was a key part of the New Product Development Process.

- Project Management: Refers to the lifecycle management of new product development efforts [6].

The structure of each maturity level is simple, the same and composed of a set of goals.

Goals are classified as generic goals and specific goals. A generic goal describes the characteristics that must be present to institutionalize the processes that implement a process area. A specific goal describes the unique characteristics that must be present to satisfy the process area [2]. Practices are expected components for satisfying goals. Practices are classified as generic practices and specific practices. A generic practice is the description of an activity that is considered important in achieving the associated generic goal. A specific practice is the description of an activity that is considered important in achieving the associated specific goal [2].

The following figure describes the global architecture of a maturity level of the staged representation.

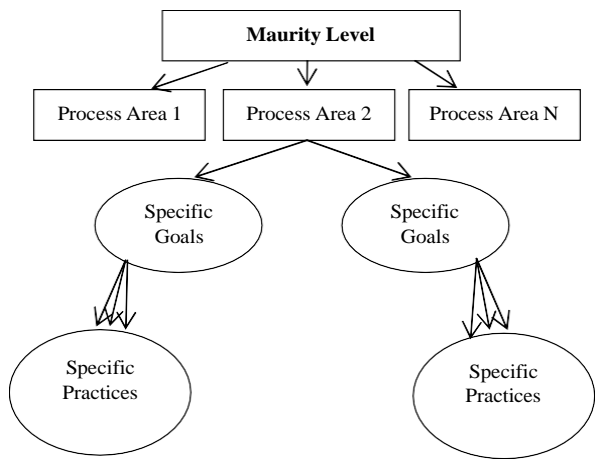

Figure 1: Maturity level Architecture for Staged Representation

## 3.2 PMBOK

The PMBOK (Project Management Body of Knowledge) is the official guide published by the Project Management Institute (PMI) that promotes a common vocabulary for project management [7] as well as the "Generally Recognized, Good Practices" of effective project management. A "generally recognized, good practice," as defined by the PMI, is a value adding practice (skills, tools, and techniques) that enhances the chances of success and applies to most projects, most of the time [8].

The PMBOK guide is process-based; it describes work as being accomplished by 42 overlapping and interacting processes divided into five process groups and nine knowledge areas (chapters). In other words, each of the nine knowledge areas contains the processes that need to be accomplished within its discipline in order to achieve an effective project management program. Similarly, each of these processes also falls into one of the five basic process groups, creating a matrix structure such that every process can be related to one knowledge area and one process group [7]. A PMBOK process is a set of Inputs, Techniques and tools and Outpus as shown in [Figure 2].

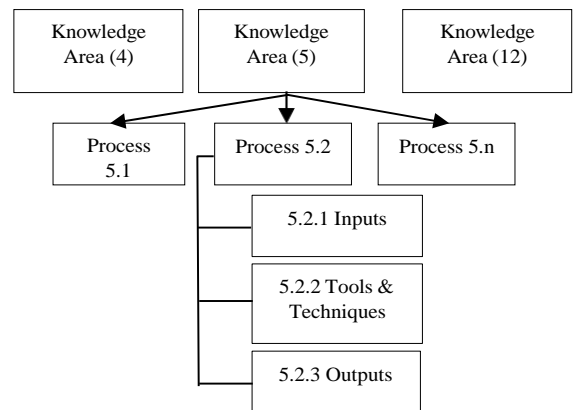

Figure 2: PMBOK Architecture

Project management processes can be divided into five groups, each containing one or more processes:

- The initiating process refers to project kick-off and initial scope definition used at the beginning of a project or a phase if desired. The goals of the upcoming work should be understood.

- The planning process refers to project scope definition and planning. It describes necessary activities to attain the project objectives. Budget and schedule should be estimated and project execution planed.

- The executing process refers to project's work executing done to satisfy the project specifications. It includes tools for quality assurance, communicating progress, and developing and managing the project team, stakeholders and sponsors.

- The Monitoring and Controlling Process refers to progress and performance monitoring and content and quality reviewing. The project objectives should been met, and project cost, schedule, quality, risk, and scope should been controlled.

- The Closing Process refers to all activities across all process groups to gain customer acceptance of the project and formally close the project.

# 4. Mapping CMMI to PMBOK

## 4.1 High level mapping between CMMI and PMBOK

Before starting a detailed comparison, it would be wise to understand the organizational and architectural features of the two frameworks. The following table presents high level comparison between CMMI and PMBOK. It is based partially on a previous study, published on 2007 by the software Engineering Institute (SEI) [9], that we revised in order to introduce characteristics of the new editions of CMMI and PMBOK.

Table 3: High level comparison between CMMI and PMBOK

| PMBOK | CMMI |
|---|---|
| A set of 42 processes organized into 5 project management process groups and 9 knowledge areas. | CMMI process areas organized into specific goals composed to specific practices, and generic goals composed to generic practices. |
| PMBOK specifies possible interfaces between process groups and their overlap across project timeline | CMMI does not endorse any particular approach to development |
| Processes are defined with inputs, tools and techniques, outputs. Interfaces between processes have been defined. | Specific and generic practices define CMMI requirements, and provide liberty for interpretation and design of process solution to meet these requirements. |
| PMBOK focuses only on Project Management activities. | CMMI focuses on Planning, Resourcing Monitoring and Control, Senior management reporting for all Project Management, Engineering, and Support process areas through Generic Practices. |
| PMBOK is open to all disciplines; it does not address specific project types. | Addresses larger organizations composed of engineering projects. |
| PMBOK supports training and Project Managers for PMP Certifications. | CMMI supports organizations to improve processes for achievement of maturity/ capability models. |
| PMBOK uses common terminologies across Project Management community. | CMMI does not prescribe any specific terminology to be followed. It is up to practitioners to adopt some terminology and satisfy framework expectations. |
| PMBOK dresses Project Management activities without focusing on Agile Approaches. | CMMI contains notes about the using of Agile Approaches on selected PAs. This can help those who use Agile methods to interpret CMMI practices in their environments |
| The use of PMBOK processes depends on organization needs. There is no formal appraisal for PMBOK compliance. | Organization need to satisfy CMMI Maturity level requirements (described in GPs and SPs) through process solutions, and ensure implementation of process solution. It needs to be formally appraised by SEI (SCAMPI A appraisal). |

## 4.2 Detailed mapping of CMMI to PMBOK

Because we cannot show the complete mapping in this paper, we had to take some special consideration to summarize this study as shown in [Table 4]. First, for Tools and Techniques elements of PMBOK, CMMI practices describe what to do, but not how to do it or who should do it. That is why specific practices do not give techniques or methods to implement PAs in CMMI. In our study, we consider Tools are provided by GP 2.3 (Provide Resources), and Techniques are defined by GP 2.2 (Plan the Process). Second, Inputs in PMBOK are generally produced as Outputs of others PMBOK processes. Implicitly, most Inputs are mapped based on Outputs mapping to CMMI practices. Moreover, for any mapping, it would be wise to bring the two models based on the same level of granularity. In our case, as defined before, CMMI practices describe the expected behavior of an organization or project to improve their processes. Based on this concept, and in order to keep the same granularity level of mapping, we make a mapping of CMMI PA to the objectives and the Outputs elements of each PMBOK process.

For a better understanding, we used the same presentation processes in PMBOK. In the first column [Table 4], each process is shown below its PMBOK chapter. While, we present, in the second column CMMI Process Areas.

Table 4: Mapping CMMI to PMBOK

| PMBOK | CMMI |
|---|---|
| **4. Project Integration Management** | |
| 4.1 Develop Project Charter | **IPM, GP2.1** |
| 4.2 Develop Project Management Plan | **PP, IPM, REQM, CM, GP2.1, 2.2, 2.3, 2.6** |
| 4.3 Direct and Manage Project Execution | **REQM, IPM, CM, PP, IPM, GP2.1, 2.6** |
| 4.4 Monitor and Control Project Work | **PMC, MA, GP2.8** |
| 4.5 Perform Integrated Change Control | **PMC, CM, IPM, GP2.10** |
| 4.6 Close Project or Phase | **IPM, GP3.2** |
| **5. Project Scope Management** | |
| 5.1 Collect Requirements | **REQM** |
| 5.2 Define Scope | **RD** |
| 5.3 Create WBS | **PP** |
| 5.4 Verify Scope | **RD** |
| 5.5 Control Scope | **PMC** |

| 6. Project Time Management | |
|---|---|
| 6.1 Define Activities | **PP, IPM** |
| 6.2 Sequence Activities | **PP, IPM** |
| 6.3 Estimate Activity Resource | **PP, IPM, GP2.3** |
| 6.4 Estimate Activity Durations | **PP, IPM** |
| 6.5 Develop Schedule | **PP, IPM** |
| 6.6 Control Schedule | **PMC** |
| **7. Project Cost Management** | |
| 7.1 Estimate Costs | **PP, GP 2.3** |
| 7.2 Determine Budget | **PP** |
| 7.3 Control Costs | **PMC** |
| **8. Project Quality Management** | |
| 8.1 Plan Quality | **PPQA, RD, MA, QPM, VER, VAL, OPP** |
| 8.2 Perform Quality Assurance | **PPQA, RD, MA, QPM, VER, VAL, GP2.6,2.9** |
| 8.3 Perform Quality Control | **PPQA, PMC, PMC, MA, CM, VER, VAL, OPP, GP2.8, 2.9, 2.10** |
| **9. Project Human Resource Management** | |
| 9.1 Develop HR plan | **IPM,PP,OT, GP2.2** |
| 9.2 Acquire project Team | **IPM** |
| 9.3 Develop Project Team | **IPM, OT, GP2.5, 2.7** |
| 9.4 Manage Project Team | **IPM, GP2.7** |
| **10. Project Communications Management** | |
| 10.1 Identify Stakeholders | **PP, REQM, GP2.4, 2.7** |
| 10.2 Plan Communications | **IPM, REQM** |
| 10.3 Distribute Information | **IPM, GP2.7** |
| 10.4 Manage Stakeholder expectations | **IPM, REQM, GP2.7** |
| 10.5 Report Performance | **PMC** |

| 11. Project Risk Management | |
|---|---|
| 11.1 Plan Risk Management | **RSKM, GP2.2** |
| 11.2 Identify Risks | **RSKM, PP** |
| 11.3 Perform Quantitative Risk Analysis | **RSKM**, PP |
| 11.4 Perform Qualitative Risk Analysis | **RSKM, PP** |
| 11.5 Plan Risk Responses | **RSKM** |
| 11.6 Monitor and Control Risks | **RSKM, PMC** |
| **12. Project Procurement Management** | |
| 12.1 Plan Procurement | **SAM, GP2.3** |
| 12.2 Conduct Procurements | **SAM** |
| 12.3 Administer Procurements | **SAM** |
| 12.4 Close Procurements | **IPM, GP3.2** |

## 5. Mapping analysis of CMMI and PMBOK

5.1 Similarities between CMMI and PMBOK

Processes addressed by both frameworks:

- *Requirements Management or Scope Control:* The objective is to focuses on understanding customer needs, translate and document them into measured requirements. Cost, schedule and quality planning are all based on these requirements which should be detailed and elicited enough to be used in the implementation and monitoring of project activities.

- *Project Planning*: Project planning focuses on establishing realistic estimates for the work to be performed by developing a project plan and obtaining commitments to this Plan. Planning will explore all aspects of scope, time, costs, quality, communication, risk and procurements.

- *Managing and Controlling Project Execution:* the project manager should establish sufficient visibility of project performances. Deviations from the plan should be detected and corrected. Corrections may include re-planning the work or taking actions that will allow the team to run the plan.

- *Quality assurance:* The purpose of quality assurance is to provide visibility of the used processes and the manufactured products by reviewing these products and processes to ensure compliance with the established standards.

- *Supplier or procurement management*: The aim is to manage the acquisition of products from suppliers, managing procurement relationship and monitoring contract performance.

- *Risk Management*: The objective is to identify, analyze and eliminate risks; they must be identified and documented, evaluated, and categorized according to their potential impact. Planning risk-handling activities to mitigate adverse impacts on achieving objectives

- *Measurement and Quantitative Project Management:* CMMI and PMBOK define a set of measurements that must be followed to ensure that the project is clearly defined, managed, and completed.

5.2 How PMBOK supplements CMMI?

- *Project charter:* The objective of the process "*Develop project charter*" is to provide a document that formally authorizes the initiation of the project and gives authority to the project Manager. The approved charter is the project's trigger and the sponsor is the defendant of the project and the appropriate of funding project.
  PMBOK gives details of developing the project charter, tools and techniques which can be used, and the charter's contents.

- *Planning :* PMBOK provides more detail for the planning; it presents additional planning documents such as:
  - o Scope Management plan,
  - o Schedule Management plan,
  - o Cost Management plan,
  - o Human resource plan,
  - o Communication Management plan
  - o Procurement Management Plan.

  The PMBOK provides more guidance and details for Time Management (define activities, sequence activities, estimate activity resources, estimate activity durations, develop and control schedule). It gives also several tools and techniques for Time Management.

- *Management and Control:* For Monitoring, the PMBOK provides formulas and techniques of performance measurement and forecasting using earned value.

- *Human Resources Management:* This Process Area covers all processes relating to planning, acquisition; developing and managing project team.

- *Quality Management*: The PMBOK consider the cost of quality and provides tools with descriptions of quality planning: Design of Experiments, Cost-Benefit Analysis and Benchmarking. The PMBOK provides tools with descriptions of quality control: Cause and effect diagram, control charts, Flowcharting, histogram, Pareto chart, chart run, scatter diagram, statistical sampling and Inspection.

- *Risk Management*: The Risk Management in PMBOK gives more details on how to plan and budget risks, gives examples of risk parameters, how to identify risks, how to perform qualitative and quantitative analysis and how to plan risk responses.

- *Procurement Management:* The project Procurement Management details how to plan, conduct, and administer procurements. The organization can be either the buyer or seller of the products or services. It includes the contract and purchase orders management

- *Close Project:* The PMBOK gives more details for administrative and contracts closure and how a formal acceptance of the product should be done.

5.3 How CMMI complements PMBOK?

- *Engineering Best Practices :* The Engineering Best Practices consist of eliciting customer needs, developing customer and product requirements, design, develop, and implement solutions to requirements, ensuring requirements traceability. The CMMI engineering involves assembling the product and managing the internal and external interface planning and preparing the integration, verification and validation of the final product.

- *Using Agile Approaches:* To help organizations who use agile methods to interpret CMMI practices in their environments, notes have been added to selected process areas. These notes are added to the following

PAs in CMMI-DEV: CM, PI, PMC, PP, PPQA, RD, REQM, RSKM, TS, and VER. Such approaches are characterized by the following [10]:
- Direct involvement of the customer in product development
- Use of multiple development iterations to learn about and evolve the product
- Customer willingness to share in the responsibility for decisions and risk

- *Organizational Process Definition*: CMMI focuses heavily on the process. It provides a specific intent to establish and maintain a usable set of organizational process assets, work environment standards, and rules and guidelines for teams. The organization's "set of standard processes" is tailored by projects to create their defined processes. Other organizational process assets are used to support tailoring and implementing defined processes [2].

- *Organizational Process Management* : The Organizational process management aims to:
  - Establish and maintain a usable set of organizational process assets and work environment standards
  - Develop people's knowledge to perform their roles effectively and efficiently in the organization.
  - Establish quantitative understanding standards to support quality and performance objectives.
  - Deploy incremental and innovative improvements that measurably improve the organization's processes and technologies.

- *Configuration management*: CMMI provides more guidance to establish and maintain the integrity of work products using configuration identification.
  It involves the following activities:
  - Identifying the configuration of selected work products that compose baselines at given points in time
  - Controlling changes to configuration items
  - Building or providing specifications to build work products from the configuration management system
  - Maintaining the integrity of baselines
  - Providing accurate status and current configuration data to developers, end users, and customers

- *Decision Analysis*: Decision Analysis process is a formal method of evaluating possible decisions and proposing solutions.

- *Continuous processes improvement:* CMMI provides capability levels. Each process undergoes gradual adjustments from one capability level to another increasing its performances. Each capability level builds the foundation needed to acquire the higher level. Especially, the level 5 focuses on continuous improvement of process performance based on incremental and innovative ameliorations.
  CMMI use the data to make decisions concerning the definition of processes, both at the project and during the change process in the organization.

## 6. Conclusion and future work

CMMI and PMBOK are not only compatible but also complementary. The PMBOK fits perfectly into a continuous improvement processes approach based on CMMI framework by offering proven practices from the experience of thousands of people working in the PMI all over the word. As we know, companies are currently looking for the efficiency of their investments before any other consideration. This could not be possible without the control of its projects to ensure the good financial health of business and the high level of customer satisfaction. The combination of both PMBOK and CMMI presents more efficient and complete framework for an organization that seeks to improve in depth and continuously its way of working.

It is nice to have available a repository of good practices to deploy. However, the deployment is not easy as it appears. This leads us to ask some questions: How PMBOK techniques and tools can be used to dress CMMI processes? Under what conditions and according to which criteria we can deploy these new approach (Organization types, methodology…)? What are the key factors of success of such as approach?...

On other hand, as we previously present in this study, CMMI (v1.3) practices have been enriched by additional notes to support organizations working in agile context. So, in what extent it is possible to combine CMMI and PMBOK with the agile manifesto?

In future research, we try to respond of all of this questions, and plan to conduct experiments and studies to confirm how effective can we deploy PMBOK and CMMI together.